\documentclass[sigconf]{acmart}

\usepackage{tabularx,booktabs}

\usepackage{amsmath}

\usepackage{array}
\usepackage{algorithm}
\usepackage{algorithmic}
\usepackage{url}
\usepackage{float}  
\usepackage{subfigure}
\usepackage{multirow}  
\usepackage{enumitem}
\usepackage[bottom]{footmisc}
\usepackage{makecell}    

\AtBeginDocument{%

\setlength{\abovedisplayskip}{1pt}
\setlength{\belowdisplayskip}{1pt}
\setlength{\abovedisplayshortskip}{1pt}
\setlength{\belowdisplayshortskip}{1pt}
}

\setcopyright{acmlicensed}
\copyrightyear{2018}
\acmYear{2018}
\acmDOI{XXXXXXX.XXXXXXX}
\acmConference[Conference acronym 'XX]{Make sure to enter the correct
  conference title from your rights confirmation email}{June 03--05,
  2018}{Woodstock, NY}
\acmISBN{978-1-4503-XXXX-X/2018/06}

\title{MaRI: Accelerating Ranking Model Inference via Structural Re-parameterization in Large Scale Recommendation System}
\author{Yusheng Huang, Pengbo Xu, Shen Wang, Changxin Lao, Jiangxia Cao$^\star$, Shuang Wen, Shuang Yang, Zhaojie Liu, Han Li, Kun Gai}
\thanks{$^\star$ Jiangxia Cao is the corresponding author.} 
\affiliation{
  \institution{Kuaishou Technology, Beijing, China}
  \country{\{huangyusheng, xupengbo03, wangshen, laochangxin, caojiangxia, wenshuang, yangshuang08, zhaotianxing, lihan08\}@kuaishou.com}\country{kun.gai@qq.com}
}

\begin{teaserfigure}
  \includegraphics[width=0.99\textwidth]{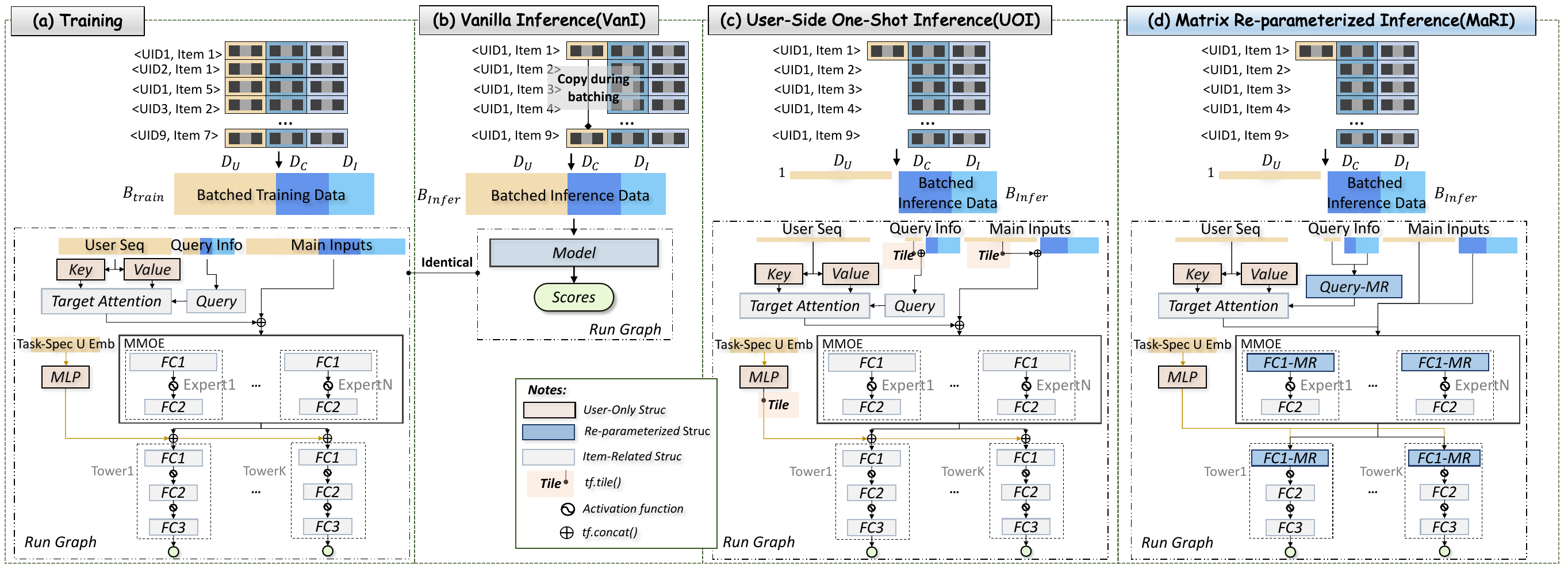}
  \caption{The proposed Matrix Re-parameterized Inference (MaRI) vs. Baseline Paradigms (Vanilla, User-Side One-Shot)}
  \label{fig:main} 
\end{teaserfigure}

\renewcommand\footnotetextcopyrightpermission[1]{}

\begin{document}

\renewcommand{\shortauthors}{Huang et al.}

\begin{abstract}
  Ranking models, i.e., coarse-ranking and fine-ranking models, serve as core components in large-scale recommendation systems, responsible for scoring massive item candidates based on user preferences. 
  To meet the stringent latency requirements of online serving, structural lightweighting or knowledge distillation techniques are commonly employed for ranking model acceleration.
  However, these approaches typically lead to a non-negligible drop in accuracy.
  Notably, the angle of lossless acceleration by optimizing feature fusion matrix multiplication, particularly through structural reparameterization, remains underexplored. 
  In this paper, we propose MaRI, a novel \underline{Ma}trix \underline{R}e-parameterized \underline{I}nference framework, which serves as a complementary approach to existing techniques while accelerating ranking model inference without any accuracy loss. 
  MaRI is motivated by the observation that user-side computation is redundant in feature fusion matrix multiplication, and we therefore adopt the philosophy of structural reparameterization to alleviate such redundancy. 
  Furthermore, we propose a graph coloring algorithm (GCA) to automate the implementation of MaRI.
  Besides, a key insight is that a neat input layout is critical to MaRI’s efficiency, while a fragmented layout causes a performance degradation of nearly 38\%.
  Online A/B test conducted on a real-world large-scale recommendation system demonstrate that MaRI achieves a speedup of up to \textbf{1.3}$\times$ with no loss in accuracy and a \textbf{5.9\%} coarse-ranking resource reduction.
\end{abstract}




\maketitle

\section{Introduction}
Short video platforms such as Kuaishou\footnote{www.kuaishou.com}, have witnessed rapid growth in recent years. To improve user experience, sophisticated recommendation systems (RS) have been built to capture users' long- and short- term interests, processing an enormous number of requests efficiently.

Modern industrial RS follow a multi-stage recommendation pipeline:\footnote{Recent advances in generative RS \cite{deng2025onerec,zhou2025onerec,guo2025onesug,wei2025oneloc} may adopt an end-to-end pipeline, which however falls outside the scope of this work.} 
(1) Retrieval, which quickly matches user interests with relevant items from a large candidate pool. 
(2) Coarse-ranking, which efficiently scores thousands of candidates under tight latency constraints using a lightweight deep neural network (DNN)
(3) Fine-ranking, which makes precise predictions for a small subset of items ($\sim100$) using a complex DNN model.

\begin{figure}[t!]
  \centering
  \includegraphics[width=0.75\linewidth]{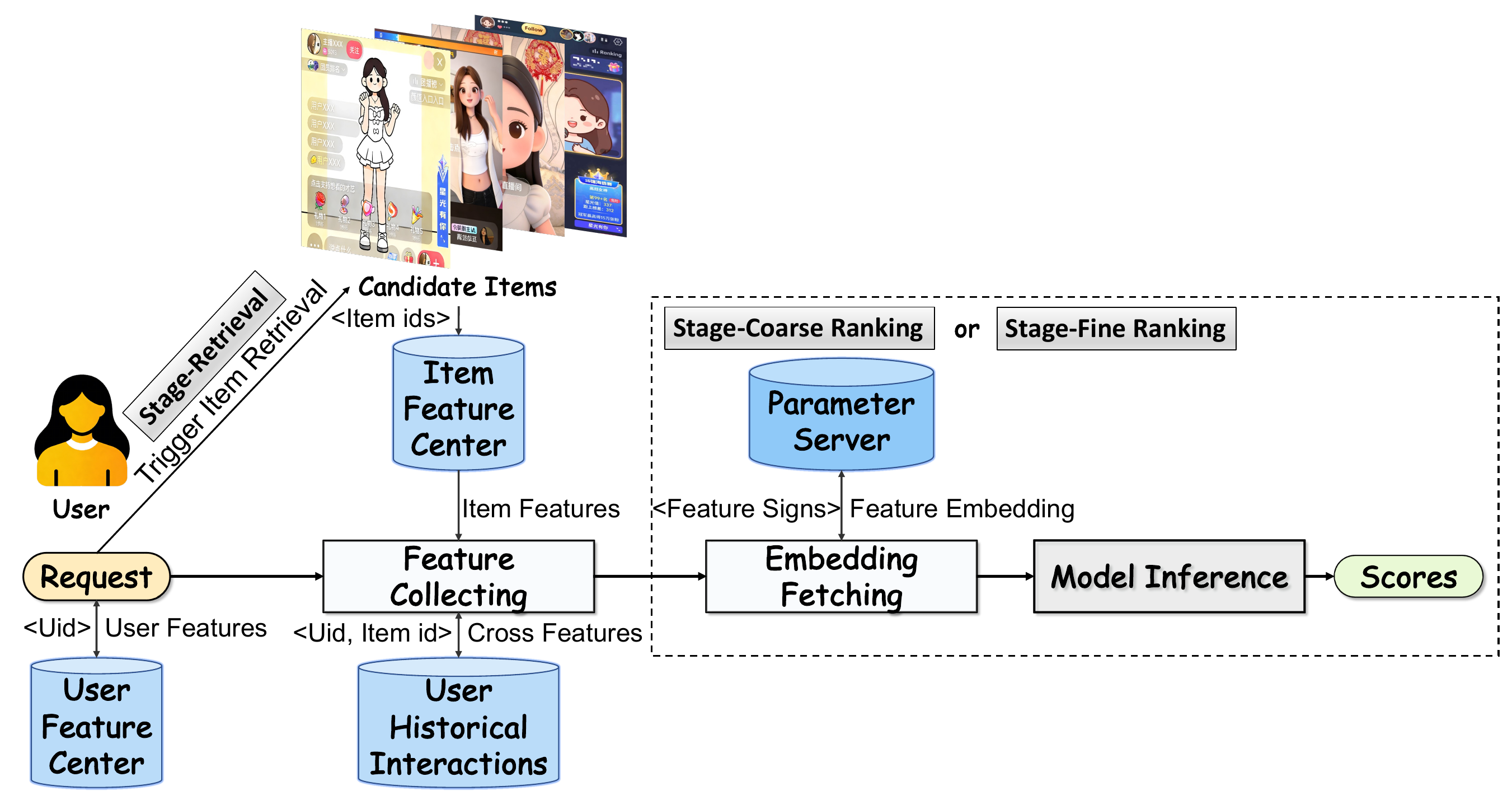}
  \caption{Inference Workflow of Our Online Large-Scale Recommendation System.}
  \label{fig:overall_infer}
\vspace{-0.6cm}
\end{figure}

Figure \ref{fig:overall_infer} shows the key inference stages related to ranking models triggered by a user request: feature collection, embedding fetching, and model inference (the most latency consuming stage). 
Hence, balancing ranking model's inference efficiency and prediction accuracy is a critical topic in industrial RS. 
Existing acceleration solutions mainly fall into three categories: lightweight model design, knowledge distillation, and model–engine co-optimization.\footnote{Detailed related works are presented in Appendix~\ref{app:related_works}.}

Besides the aforementioned related methods in Appendix~\ref{app:related_works}, we identify a promising while not well-explored optimization direction: the acceleration of matrix multiplication (MatMul) for feature fusion in ranking models. 
For modeling high-order feature interactions, user, item\&cross features are concatenated and fed into DNNs and their variants,
where MatMul serves as the primary means of feature fusion.
As fine-grained features continue to proliferate in industrial recommendation scenarios, the dimension of the concatenated feature vectors surges, rendering MatMul the dominant contributor to inference latency. 

Current MatMul acceleration solutions are mostly focused on hardware and system-level optimizations \cite{Kerr2017cutlass,chen2018tvm, NVIDIAndTensorRT}, such as kernel optimization\&fusion, precision quantization and memory access optimization. 
These methods rely on specific hardware architectures or engineering implementations to improve MatMul efficiency, yet do not explicitly target the structural design of MatMul used for feature fusion in ranking models.


In this paper, we present a \textbf{model-design-level} MatMul acceleration method built upon the philosophy of \textbf{structural re-parameterization}\cite{ding2021repvgg,ding2022repmlpnet}.
Our approach keeps the model training pipeline unchanged, and re-parameterizes the feature fusion MatMul operations by decoupling user, item and cross feature computations during ranking model inference. 
Such reformulation is \textbf{mathematically equivalent}, thus achieving higher inference efficiency while maintaining lossless accuracy. 
Being hardware-agnostic, our method offers a novel complementary angle for research on efficient inference in RS.


Our contributions are summarized as follows:
\begin{itemize}[leftmargin=*,align=left]
\item We propose \textbf{MaRI} (\underline{Ma}trix \underline{R}e-parameterized \underline{I}nference), a model-design-level inference acceleration scheme for ranking models.
\item We further design \textbf{GCA} (\underline{G}raph \underline{C}oloring \underline{A}lgorithm), which automatically locates eligible MatMul operators for MaRI from model's computation graph, replacing manual search, thus enhancing the applicability of the overall framework.
\item A key insight is that a neat input layout is critical, whereas a fragmented layout leads to nearly 38\% performance degradation, which necessitates feature-and-parameter reorganization.
\item Extensive offline experiments and online deployment validate the effectiveness of the MaRI framework. In online A/B test, it maintains lossless core A/B metrics, yields up to \textbf{1.3$\times$} inference speedup on the coarse-ranking model, reduces stage latency by $2.2\%$, and cuts inference resource consumption of the coarse-ranking model by $5.9\%$. 
\end{itemize}

\section{Methodology}

\subsection{Preliminaries}

\paragraph{Ranking model.}
In RS, the core function of a ranking model is to score an incoming user against a set of candidate items. 
These scores typically represent the predicted probabilities of user's implicit feedback, such as click-through rate, long-view rate, etc. 
Generally, industrial recommendation models are well-designed DNNs, equipped with sophisticated components, including sequence modeling modules (e.g., DIN\cite{zhou2018deep}, DIEN\cite{zhou2019deep} or TWIN\cite{chang2023twin}) to capture user historical preferences, multi-task learning modules (e.g., MMOE\cite{ma2018modeling} or PLE\cite{tang2020progressive}) to handle multiple recommendation objectives, and feature crossing modules (e.g., DeepFM\cite{guo2017deepfm}, CAN\cite{zhou2020can}, Pepnet\cite{chang2023pepnet} or Rankmixer\cite{zhu2025rankmixer}) to extract user-item co-action patterns. 
(Fig.~\ref{fig:main} presents a simplified ranking model.)
It takes as input a rich feature set and outputs the predicted scores for each given user-item pair. 
The input features usually encompass user-side features (e.g., user ID, historical interaction sequences) represented by $X_{user} \in R^{1\times D_{u}}$ where $1$ denotes per single user and $D_{u}$ is the embedding dimension, item-side features (e.g., item ID, item attributes) represented by $X_{item} \in R^{B\times D_{i}}$ where $B$ denotes the size of candidate items and $D_{i}$ is the embedding dimension, and user-item interaction features (e.g., the interacting frequency between user and item) represented by $X_{cross} \in R^{B\times D_{c}}$. 
During ranking model training (cf. Fig.~\ref{fig:main}(a)), interaction logs across different users are grouped to form the training dataset. 
For inference, Fig.~\ref{fig:main}(b) and Fig.~\ref{fig:main}(c) depicts two commonly used inference methods: vanilla inference (VanI) and user-side one-shot inference (UOI).

\paragraph{Vanilla Inference (VanI)}
In industrial deployment, each inference batch typically contains only one user request due to massive candidate items, and over-sized item pools are further split into mini-batches.
In VanI (cf. Fig~\ref{fig:main}(b)), item and cross features of the target user’s candidates are batched together, while user features are replicated and broadcast to match the batch dimension $B$ to form valid input.
This design ensures the inference computation graph behaves identically to that in training, but suffers from critical user-side redundant computation.
For instance, the user interaction sequence \( X_{u\_s} \in \mathbb{R}^{1\times L_{u\_s}\times D_{u\_s}} \) is replicated $B$ times as \( X^{Tiled}_{u\_s} \), causing the key and value projections to be redundantly computed B times during cross-attention:
\begin{equation}
  \label{eq:cross_attention}
  E_{\langle\text{i\_q}, u^{Tiled}_s\rangle} = \text{Cro\_Att}\bigl(
     \text{q=}X_{\text{i\_q}}W_Q, \text{k=}X^{Tiled}_{\text{u\_s}}W_K, \text{v=}
      X^{Tiled}_{\text{u\_s}}W_V
  \bigr),
  \end{equation}
  where \( E_{\langle\text{i\_q}, u^{Tiled}_s\rangle} \) denotes the cross-attention output between candidate items and the replicated user sequence. \( W_K, W_V \in \mathbb{R}^{D_{u\_s}} \) and \( W_Q \in \mathbb{R}^{D_{i\_q}} \) are the projection matrices for key, value, and query, respectively.

\paragraph{User-Side One-Shot Inference (UOI)} \footnote{Also known as ``User Compressed Inference'' at Kuaishou, this is a mature, widely deployed industrial paradigm.}
To address the key limitation of VanI, UOI computes the user-only module once without redundant batch-wise calculation by avoiding user feature replication during batching.\footnote{Detailed FLOPs analysis is in Appendix~\ref{app:flops_uoi}.}
User tensor replication is deferred to just before concatenation with item/cross features (cf.``Tile'' in Fig.~\ref{fig:main}(c)). 




\subsection{Matrix Re-parameterization for Ranking Models}\label{sec:2_repara}

While UOI effectively eliminates part of the user-side redundant computation in VanI, redundant computation still arises in MLP and other matrix-multiplication layers, after user features are tiled, concatenated with item/cross features, and fed into these layers. In our proposed MaRI, we aim to further resolve this issue.



Preliminary: Consider a MatMul operation, where $\mathbf{A} = \left[\mathbf{A}_1,\ \dots,\ \mathbf{A}_I\right] \in \mathbb{R}^{n \times k}$ with $\mathbf{A}_i \in \mathbb{R}^{n \times k_i}$ being the $i$-th block of the column-partitioned matrix $\mathbf{A}$, and where $\mathbf{B} = \begin{bmatrix}\mathbf{B}_1^T \\ \vdots \\ \mathbf{B}_I^T\end{bmatrix} \in \mathbb{R}^{k \times m}$ with $\mathbf{B}_i^T \in \mathbb{R}^{k_i \times m}$ denoting the $i$-th block of the row-partitioned matrix $\mathbf{B}$. We have
\begin{equation}
    \mathbf{AB} = \sum_{i=1}^I \mathbf{A}_i \cdot \mathbf{B}_i^T.
    \label{eq:prel}
\end{equation}

For the MatMul layer in ranking models, let $\mathbf{W} \in \mathbb{R}^{D \times d}$ denote the weight matrix, where $D = D_{\text{user}} + D_{\text{item}} + D_{\text{cross}}$ is the concatenated feature dimension, and $d$ is the output dimension.
We could partition $\mathbf{W}$ into three row blocks corresponding to user, item, and cross features respectively:
\begin{equation}
\mathbf{W} = \begin{bmatrix}
\mathbf{W}_{\text{user}} \\
\mathbf{W}_{\text{item}} \\
\mathbf{W}_{\text{cross}}
\end{bmatrix},
\label{eq:W}
\end{equation}
with $\mathbf{W}_{\text{user}} \in \mathbb{R}^{D_{\text{user}} \times d}$, $\mathbf{W}_{\text{item}} \in \mathbb{R}^{D_{\text{item}} \times d}$, $\mathbf{W}_{\text{cross}} \in \mathbb{R}^{D_{\text{cross}} \times d}$.

Let the tiled user feature, item feature, and cross feature be concatenated as:
\begin{equation}
\mathbf{X^{\text{U\_Tiled}}} = \bigl[\,\mathbf{X}_{\text{user}}^{\text{Tiled}},\ \mathbf{X}_{\text{item}},\ \mathbf{X}_{\text{cross}}\,\bigr] \in \mathbb{R}^{B \times D},
\label{eq:X}
\end{equation}
then MatMul in the ranking model is defined as
\begin{equation}
    \text{MatMul}(\mathbf{X^{\text{U\_Tiled}}}, \mathbf{W}) := \mathbf{X^{\text{U\_Tiled}}}\mathbf{W}.
    \label{eq:ori_matmul}
\end{equation}

By the block matrix multiplication rule in (\ref{eq:prel}), the forward MatMul can be decomposed into a sum of three separate terms\footnote{$\mathbf{X}_{\text{user}}$ is tiled to $\mathbf{X}_{\text{user}}^{\text{Tiled}}$ during data batching in VanI and just prior to user-item/cross feature concatenation in UOI.}:
\begin{equation}
    \mathbf{X^{\text{U\_Tiled}}}\mathbf{W}
    = \mathbf{X}_{\text{user}}^{\text{Tiled}} \mathbf{W}_{\text{user}}
    + \mathbf{X}_{\text{item}} \mathbf{W}_{\text{item}}
    + \mathbf{X}_{\text{cross}} \mathbf{W}_{\text{cross}}.
    \label{eq:matmul_decompose}
\end{equation} It can be observed that $\mathbf{X}_{\text{user}}^{\text{Tiled}} \mathbf{W}_{\text{user}}$ is \textbf{repeatedly computed across the batch dimension}, leading to substantial redundant computation.

To address this issuse in MaRI, we propose the structural re-parameterizated $\text{MatMul}_{\text{MaRI}}$, by further defering the replication of $\mathbf{X}_{\text{user}}$ and re-parameterizing the MatMul operation (i.e. Eq.~\ref{eq:ori_matmul}) into:
\begin{equation}
  \begin{aligned}
    \text{MatMul}_{\text{MaRI}}(\mathbf{X}, \mathbf{W}) & := \text{Tile}(\mathbf{X_{\text{user}}}\mathbf{W_{\text{user}}}, B) \\
      & + \mathbf{X_{\text{item}}}\mathbf{W_{\text{item}}} + \mathbf{X_{\text{cross}}}\mathbf{W_{\text{cross}}},
  \end{aligned}
    \label{eq:re_matmul}
\end{equation}
where $\text{Tile}(, B)$ represents tiling the tensor along the batch dimension. 
The re-parameterization in Eq.~\ref{eq:re_matmul} reduces FLOPs by $\approx D_u/(D_u+D_i+D_c)$ (for $B \gg 1$) via removing redundant user-side computation.\footnote{Full FLOPs analysis is in Appendix \ref{app:flops_mari}.}
Another benefit is that item-side computation is decoupled and could be further offloaded to a near-line computation runner.\footnote{There exists a trade-off between offloaded item-side computation and bandwidth consumption; we do not implement this scheme as our system reaches the bandwidth constrain.}

\subsection{Graph Coloring Algorithm for Eligible Node Detection}\label{sec:2_gca}

In industrial ranking model with intricate structures, manually identifying MatMul nodes eligible for structural re-parameterization (MaRI) is labor-intensive and error-prone. 
To address this challenge, we propose the Graph Coloring Algorithm (GCA, cf. Algorithm~\ref{alg:gca}) for automated detection of MaRI-optimizable MatMul nodes. 
GCA first initializes node colors by feature types (user-side as Yellow, item/cross-side as Blue) and propagates colors via DFS. 
It then detects ``Concat'' nodes with mixed Yellow\&Blue inputs, and finally locates all MatMul nodes reachable from these ``Concat'' nodes through non-computational paths as the potential targets for MaRI optimization.

\vspace{-5pt}
\subsection{A Bitter Lesson: Feature and Parameter Reorganization}\label{sec:2_remap}
In industrial practice, online ranking models have evolved over years of iterative development.
The input feature layout is often less neatly-structured than Eq.~\ref{eq:X}, with features interleaved arbitrarily across user, cross, and item domains, e.g.,
\[
\mathbf{X} = \bigl[\,\mathbf{X}_{\text{user}_{\text{f}1}},\ \mathbf{X}_{\text{cross}_{\text{f}1}},\ \mathbf{X}_{\text{item}_{\text{f}1}},\ \mathbf{X}_{\text{user}_{\text{f}2}},\ \mathbf{X}_{\text{item}_{\text{f}2}},\ \mathbf{X}_{\text{cross}_{\text{f}2}},\ \dots\,\bigr] \in \mathbb{R}^{B \times D}.
\]

In this scenario, directly applying $\text{MatMul}_{\text{MaRI}}$ creates numerous small, fragmented MatMuls, leading to severely sub-optimal inference efficiency.
In our initial experiments, inference performance degrades by nearly \textbf{38.8\%} relative to the online UOI baseline.

Thus, to resolve this practical issue, it is necessary to reorganize the input feature layout and remap the corresponding learnable parameters into a clean, unified structure as in Eq.~\ref{eq:X},
so that $\text{MatMul}_{\text{MaRI}}$ only generates three large MatMuls.

\vspace{-6.5pt}
\subsection{Overall Architecture of MaRI}
The overall workflow of MaRI comprises three key steps:
(1) Apply the Graph Coloring Algorithm (GCA) proposed in Sec.~\ref{sec:2_gca} to automatically detect MaRI-optimizable MatMul nodes.
(2) Reorganize input features and remap corresponding learnable parameters into a neat, structured format (cf. Eq.~\ref{eq:X} and Sec.~\ref{sec:2_remap}).
(3) Replace the detected optimizable MatMul nodes with the proposed $\text{MatMul}_{\text{MaRI}}$ (cf. Sec.~\ref{sec:2_repara}) for inference acceleration.

Fig.~\ref{fig:main}(d) illustrates an example of MaRI’s application in a simplified ranking model. 
A key distinction between MaRI and UOI lies in the handling of user-side features during MatMul involving user-item/cross interactions.
In UOI, user-side features must be tiled to the $B$ dimension prior to such MatMul. This is observed in multiple components, including the first fully-connected (FC) layer of each expert in the MMOE structure, the first FC layer of the task tower, and the query component of cross-attention. 
In contrast, MaRI replaces all these MatMul with $\text{MatMul}_{\text{MaRI}}$, thereby significantly improving inference efficiency. 

Notably, initially only the first FC layer of each MMoE expert was manually identified as optimizable in our project, while our \textbf{GCA algorithm uncovered the other two}. This highlights the effectiveness and automation of our GCA algorithm in mining potential optimizable nodes.

\vspace{-2.5pt}
\section{Experiments}
\subsection{Offline Simulation}

\begin{figure}[htb!]
  \centering
  \includegraphics[width=0.95\linewidth]{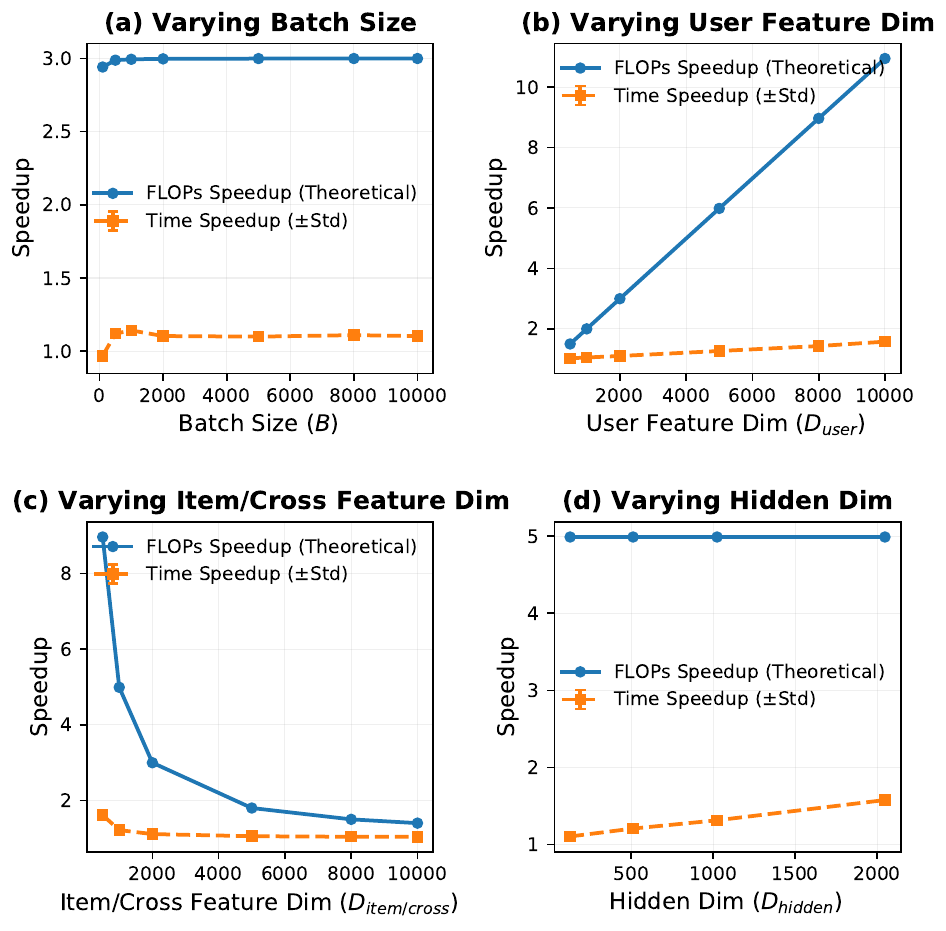}
  \caption{Offline simulation results: Performance comparison (latency/FLOPs) between $\text{MatMul}_{\text{MaRI}}$ and vanilla MatMul under varying $B$, $D_{\text{user}}, D_{\text{item/cross}}$, and $D_{\text{hidden}}$).}
  \label{fig:exp1}
\end{figure}

$\text{MatMul}_{\text{MaRI}}$ is evaluated against the vanilla MatMul under identical hardware configurations across four experiments with distinct parameter settings.\footnote{Specifically, $B \in \{100, 500, 1000, 2000, 5000, 8000, 10000\}$, $D_{\text{user}}, D_{\text{item/cross}} \in \{100, 500, 1000, 2000, 5000, 8000, 10000\}$, and $D_{\text{hidden}} \in \{64, 128, 512, 1024\}$.}
We sample a subset real-world data and report the mean $\pm$ standard deviation (std) over 100 independent runs.

Offline simulations validate the effectiveness of MaRI(cf.Figure.~\ref{fig:exp1}, and Table.~\ref{tab:exp_comprehensive} in Appendix~\ref{app:exp}): 
(1) For $B \geq 500$, runtime speedup remains $1.1\times$ indicating MaRI’s stability for large-scale candidate item batches.
(2) FLOPs and runtime speedup grow linearly with $D_{\text{user}}$ peaking at $10.95\times$ and $1.58\times$, while they decrease as $D_{\text{item/cross}}$  increases, which aligns with MaRI’s core design of optimizing redundant user-side computation.
(3) For $D_{\text{hidden}}$, MaRI’s theoretical FLOPs reduction remains constant while runtime speedup rises from $1.10\times$ to $1.58\times$ with larger hidden size.

\begin{figure}[hbt!]
  \centering
  \includegraphics[width=0.85\linewidth]{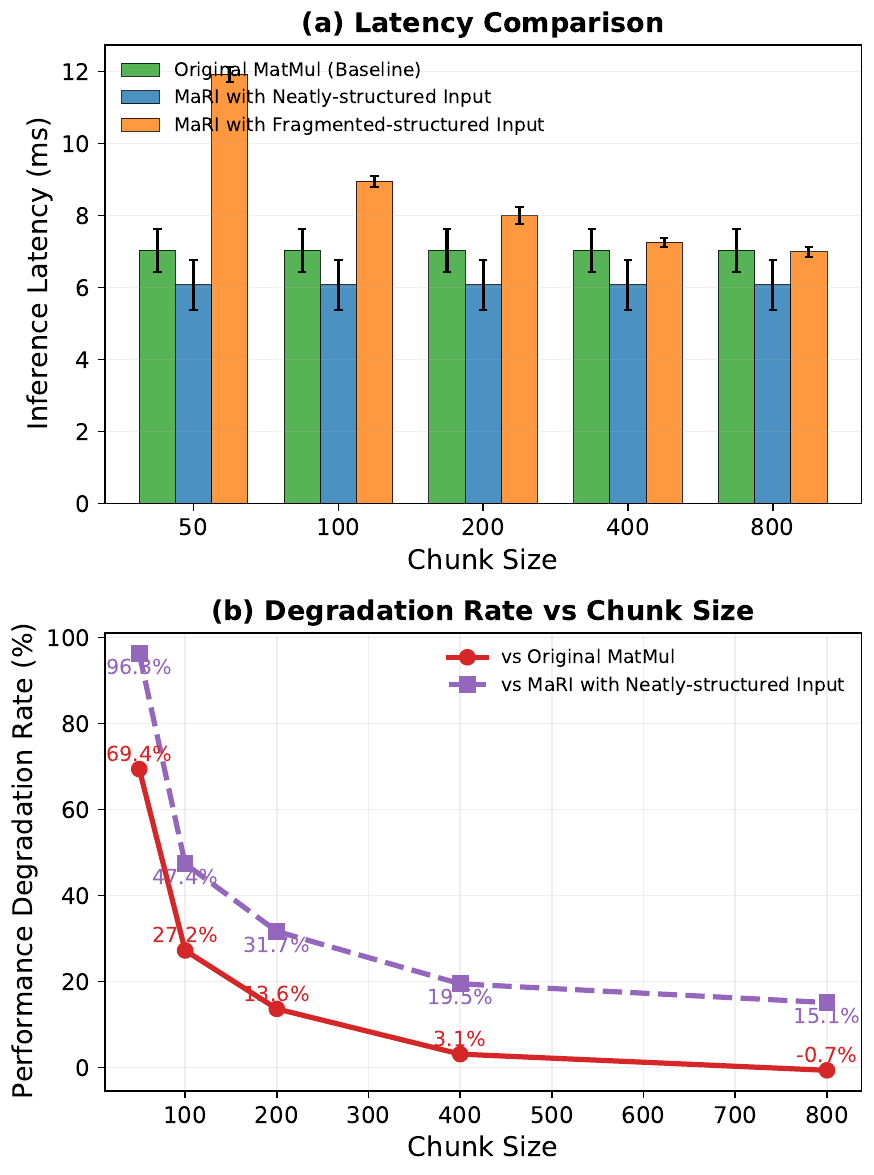}
  \caption{Fragmented MaRI vs. Vanilla MatMul and Neat MaRI.}
  \label{fig:exp2}
\end{figure}

We further decompose the neat input into chunks of different sizes to simulate fragmented feature layout in industrial scenarios.
As reported in Figure.~\ref{fig:exp2} and Table.~\ref{tab:chunk_size_performance} in Appendix~\ref{app:exp}, $\text{MatMul}_{\text{MaRI}}$ with neat input achieves the best performance. 
In contrast, \textbf{small chunk sizes lead to severe efficiency degradation}: at chunk size 50, the latency of fragmented MaRI increases by 69.4\% and 96.3\% compared with the original MatMul and the neat MaRI.
As the chunk size grows, such performance degradation is gradually alleviated but still incurs a noticeable efficiency loss of 15.1\% at chunk size 800.
These results validate the necessity of feature\&parameter re-organization for MaRI to achieve its full potential.


\subsection{Online Deployment}

We deployed our proposed MaRI method into the live-streaming recommendation systems of KuaiShou and KuaiShou Lite and carried out an online A/B test from 2025-06-21 to 2025-06-24. 
The training AUC of the MaRI optmized model remains unchanged.
Each group was allocated 5\% of the total users, where the experimental group (Exp) adopted our MaRI-optimized model, and the control group (Ctrl) employed the standard UOI method.\footnote{Both models are optimized with industrial inference engine and acceleration tools for a fair comparison.}
Online business metrics presented \textbf{no statistically significant differences} between the two groups showing that the MaRI-optimized model maintains the same level of business performance as the baseline.
As presented in Table~\ref{table:rec_pipeline_latency}, MaRI sped up model inference by \textbf{1.32$\times$(Avg)} and \textbf{1.26$\times$(P99)}, achieving a latency reduction of \textbf{2.24\%} and \textbf{2.27\%} in the coarse-ranking stage, and \textbf{0.42\%} and \textbf{0.54\%} for the entire recommendation pipeline, respectively.
Furthermore, the full deployment of MaRI on 2025-07-04 achieved a \textbf{5.9\%} hardware resource reduction for the coarse-ranking inference stage.

\begin{table}[tb!] 
  \centering
  \caption{Relative Latency Differences in Online Recommendation Pipeline (Exp vs. Ctrl).}
  \resizebox{\linewidth}{!}{
  \begin{tabular}{lcc|cc}
  \toprule[0.75pt] 
  Stage     & \multicolumn{2}{c}{\makecell{Relative Latency\\Change (\%)}} & \multicolumn{2}{c}{\makecell{Relative RunGraph\\Time Speedup}} \\
  \cmidrule(r){2-3} \cmidrule(l){4-5} 
            & KuaiShou & KuaiShou Lite & Avg & P99 \\
  \midrule
  Coarse-ranking & -2.24 & -2.27 & 1.32$\times$ & 1.26$\times$ \\ 
  Overall   & -0.42 & -0.54 & -- & -- \\ 
  \bottomrule[0.75pt] 
  \end{tabular}
  }
  \label{table:rec_pipeline_latency}
  \end{table}
\section{Conclusions}

In this paper, we propose MaRI, a MatMul acceleration framework for RS via structural reparameterization.
MaRI eliminates user-side redundant computations in user-item\&cross interaction MatMuls, and leverages a graph coloring algorithm (GCA) to automatically locate optimizable nodes.
We highlight that a neat input structure is critical to unlocking MaRI’s full acceleration potential.
Extensive offline and online experiments validate its superior effectiveness and efficiency.

\begin{acks}
The authors would like to thank the Kuaishou UniPredict Team for their valuable support in model deployment and the Kuaishou Kai Team for their assistance with model training infrastructure.
\end{acks}
\vspace{-12pt}
\section*{Presenter Bio}
Yusheng Huang is currently with Kuaishou Technology. His research interest includes generative recommendation, long sequence modeling and efficient model inference.

\vspace{-10pt}
\bibliographystyle{ACM-Reference-Format}
\bibliography{ref}


\begin{thebibliography}{29}


\ifx \showCODEN    \undefined \def \showCODEN     #1{\unskip}     \fi
\ifx \showISBNx    \undefined \def \showISBNx     #1{\unskip}     \fi
\ifx \showISBNxiii \undefined \def \showISBNxiii  #1{\unskip}     \fi
\ifx \showISSN     \undefined \def \showISSN      #1{\unskip}     \fi
\ifx \showLCCN     \undefined \def \showLCCN      #1{\unskip}     \fi
\ifx \shownote     \undefined \def \shownote      #1{#1}          \fi
\ifx \showarticletitle \undefined \def \showarticletitle #1{#1}   \fi
\ifx \showURL      \undefined \def \showURL       {\relax}        \fi
\providecommand\bibfield[2]{#2}
\providecommand\bibinfo[2]{#2}
\providecommand\natexlab[1]{#1}
\providecommand\showeprint[2][]{arXiv:#2}

\bibitem[Bin et~al\mbox{.}(2025)]%
        {bin2025real}
\bibfield{author}{\bibinfo{person}{Xingyan Bin}, \bibinfo{person}{Jianfei Cui}, \bibinfo{person}{Wujie Yan}, \bibinfo{person}{Zhichen Zhao}, \bibinfo{person}{Xintian Han}, \bibinfo{person}{Chongyang Yan}, \bibinfo{person}{Feng Zhang}, \bibinfo{person}{Xun Zhou}, \bibinfo{person}{Xiao Yang}, {and} \bibinfo{person}{Zuotao Liu}.} \bibinfo{year}{2025}\natexlab{}.
\newblock \showarticletitle{Real-time Indexing for Large-scale Recommendation by Streaming Vector Quantization Retriever}. In \bibinfo{booktitle}{\emph{Proceedings of the 31st ACM SIGKDD Conference on Knowledge Discovery and Data Mining V. 2}}. \bibinfo{pages}{4273--4283}.
\newblock


\bibitem[Chai et~al\mbox{.}(2025)]%
        {chai2025longer}
\bibfield{author}{\bibinfo{person}{Zheng Chai}, \bibinfo{person}{Qin Ren}, \bibinfo{person}{Xijun Xiao}, \bibinfo{person}{Huizhi Yang}, \bibinfo{person}{Bo Han}, \bibinfo{person}{Sijun Zhang}, \bibinfo{person}{Di Chen}, \bibinfo{person}{Hui Lu}, \bibinfo{person}{Wenlin Zhao}, \bibinfo{person}{Lele Yu}, {et~al\mbox{.}}} \bibinfo{year}{2025}\natexlab{}.
\newblock \showarticletitle{Longer: Scaling up long sequence modeling in industrial recommenders}. In \bibinfo{booktitle}{\emph{Proceedings of the Nineteenth ACM Conference on Recommender Systems}}. \bibinfo{pages}{247--256}.
\newblock


\bibitem[Chang et~al\mbox{.}(2023a)]%
        {chang2023twin}
\bibfield{author}{\bibinfo{person}{Jianxin Chang}, \bibinfo{person}{Chenbin Zhang}, \bibinfo{person}{Zhiyi Fu}, \bibinfo{person}{Xiaoxue Zang}, \bibinfo{person}{Lin Guan}, \bibinfo{person}{Jing Lu}, \bibinfo{person}{Yiqun Hui}, \bibinfo{person}{Dewei Leng}, \bibinfo{person}{Yanan Niu}, \bibinfo{person}{Yang Song}, {et~al\mbox{.}}} \bibinfo{year}{2023}\natexlab{a}.
\newblock \showarticletitle{TWIN: TWo-stage interest network for lifelong user behavior modeling in CTR prediction at kuaishou}. In \bibinfo{booktitle}{\emph{Proceedings of the 29th ACM SIGKDD Conference on Knowledge Discovery and Data Mining}}. \bibinfo{pages}{3785--3794}.
\newblock


\bibitem[Chang et~al\mbox{.}(2023b)]%
        {chang2023pepnet}
\bibfield{author}{\bibinfo{person}{Jianxin Chang}, \bibinfo{person}{Chenbin Zhang}, \bibinfo{person}{Yiqun Hui}, \bibinfo{person}{Dewei Leng}, \bibinfo{person}{Yanan Niu}, \bibinfo{person}{Yang Song}, {and} \bibinfo{person}{Kun Gai}.} \bibinfo{year}{2023}\natexlab{b}.
\newblock \showarticletitle{Pepnet: Parameter and embedding personalized network for infusing with personalized prior information}. In \bibinfo{booktitle}{\emph{Proceedings of the 29th ACM SIGKDD Conference on Knowledge Discovery and Data Mining}}. \bibinfo{pages}{3795--3804}.
\newblock


\bibitem[Chen et~al\mbox{.}(2021)]%
        {chen2021end}
\bibfield{author}{\bibinfo{person}{Qiwei Chen}, \bibinfo{person}{Changhua Pei}, \bibinfo{person}{Shanshan Lv}, \bibinfo{person}{Chao Li}, \bibinfo{person}{Junfeng Ge}, {and} \bibinfo{person}{Wenwu Ou}.} \bibinfo{year}{2021}\natexlab{}.
\newblock \showarticletitle{End-to-end user behavior retrieval in click-through rateprediction model}.
\newblock \bibinfo{journal}{\emph{arXiv preprint arXiv:2108.04468}} (\bibinfo{year}{2021}).
\newblock


\bibitem[Chen et~al\mbox{.}(2018)]%
        {chen2018tvm}
\bibfield{author}{\bibinfo{person}{Tianqi Chen}, \bibinfo{person}{Thierry Moreau}, \bibinfo{person}{Ziheng Jiang}, \bibinfo{person}{Lianmin Zheng}, \bibinfo{person}{Eddie Yan}, \bibinfo{person}{Haichen Shen}, \bibinfo{person}{Meghan Cowan}, \bibinfo{person}{Leyuan Wang}, \bibinfo{person}{Yuwei Hu}, \bibinfo{person}{Luis Ceze}, {et~al\mbox{.}}} \bibinfo{year}{2018}\natexlab{}.
\newblock \showarticletitle{$\{$TVM$\}$: An automated $\{$End-to-End$\}$ optimizing compiler for deep learning}. In \bibinfo{booktitle}{\emph{13th USENIX symposium on operating systems design and implementation (OSDI 18)}}. \bibinfo{pages}{578--594}.
\newblock


\bibitem[Corporation(nd)]%
        {NVIDIAndTensorRT}
\bibfield{author}{\bibinfo{person}{NVIDIA Corporation}.} \bibinfo{year}{n.d.}\natexlab{}.
\newblock \bibinfo{title}{NVIDIA TensorRT}.
\newblock \bibinfo{howpublished}{NVIDIA Developer Online Official Documentation}.
\newblock
\urldef\tempurl%
\url{https://developer.nvidia.com/tensorrt}
\showURL{%
\tempurl}


\bibitem[Deng et~al\mbox{.}(2025)]%
        {deng2025onerec}
\bibfield{author}{\bibinfo{person}{Jiaxin Deng}, \bibinfo{person}{Shiyao Wang}, \bibinfo{person}{Kuo Cai}, \bibinfo{person}{Lejian Ren}, \bibinfo{person}{Qigen Hu}, \bibinfo{person}{Weifeng Ding}, \bibinfo{person}{Qiang Luo}, {and} \bibinfo{person}{Guorui Zhou}.} \bibinfo{year}{2025}\natexlab{}.
\newblock \showarticletitle{Onerec: Unifying retrieve and rank with generative recommender and iterative preference alignment}.
\newblock \bibinfo{journal}{\emph{arXiv preprint arXiv:2502.18965}} (\bibinfo{year}{2025}).
\newblock


\bibitem[Ding et~al\mbox{.}(2022)]%
        {ding2022repmlpnet}
\bibfield{author}{\bibinfo{person}{Xiaohan Ding}, \bibinfo{person}{Honghao Chen}, \bibinfo{person}{Xiangyu Zhang}, \bibinfo{person}{Jungong Han}, {and} \bibinfo{person}{Guiguang Ding}.} \bibinfo{year}{2022}\natexlab{}.
\newblock \showarticletitle{Repmlpnet: Hierarchical vision mlp with re-parameterized locality}. In \bibinfo{booktitle}{\emph{Proceedings of the IEEE/CVF conference on computer vision and pattern recognition}}. \bibinfo{pages}{578--587}.
\newblock


\bibitem[Ding et~al\mbox{.}(2021)]%
        {ding2021repvgg}
\bibfield{author}{\bibinfo{person}{Xiaohan Ding}, \bibinfo{person}{Xiangyu Zhang}, \bibinfo{person}{Ningning Ma}, \bibinfo{person}{Jungong Han}, \bibinfo{person}{Guiguang Ding}, {and} \bibinfo{person}{Jian Sun}.} \bibinfo{year}{2021}\natexlab{}.
\newblock \showarticletitle{Repvgg: Making vgg-style convnets great again}. In \bibinfo{booktitle}{\emph{Proceedings of the IEEE/CVF conference on computer vision and pattern recognition}}. \bibinfo{pages}{13733--13742}.
\newblock


\bibitem[Du et~al\mbox{.}(2025)]%
        {du2025active}
\bibfield{author}{\bibinfo{person}{Yingpeng Du}, \bibinfo{person}{Zhu Sun}, \bibinfo{person}{Ziyan Wang}, \bibinfo{person}{Haoyan Chua}, \bibinfo{person}{Jie Zhang}, {and} \bibinfo{person}{Yew-Soon Ong}.} \bibinfo{year}{2025}\natexlab{}.
\newblock \showarticletitle{Active large language model-based knowledge distillation for session-based recommendation}. In \bibinfo{booktitle}{\emph{Proceedings of the AAAI Conference on Artificial Intelligence}}, Vol.~\bibinfo{volume}{39}. \bibinfo{pages}{11607--11615}.
\newblock


\bibitem[Guo et~al\mbox{.}(2017)]%
        {guo2017deepfm}
\bibfield{author}{\bibinfo{person}{Huifeng Guo}, \bibinfo{person}{Ruiming Tang}, \bibinfo{person}{Yunming Ye}, \bibinfo{person}{Zhenguo Li}, {and} \bibinfo{person}{Xiuqiang He}.} \bibinfo{year}{2017}\natexlab{}.
\newblock \showarticletitle{DeepFM: a factorization-machine based neural network for CTR prediction}.
\newblock \bibinfo{journal}{\emph{arXiv preprint arXiv:1703.04247}} (\bibinfo{year}{2017}).
\newblock


\bibitem[Guo et~al\mbox{.}(2025)]%
        {guo2025onesug}
\bibfield{author}{\bibinfo{person}{Xian Guo}, \bibinfo{person}{Ben Chen}, \bibinfo{person}{Siyuan Wang}, \bibinfo{person}{Ying Yang}, \bibinfo{person}{Chenyi Lei}, \bibinfo{person}{Yuqing Ding}, {and} \bibinfo{person}{Han Li}.} \bibinfo{year}{2025}\natexlab{}.
\newblock \showarticletitle{OneSug: The Unified End-to-End Generative Framework for E-commerce Query Suggestion}.
\newblock \bibinfo{journal}{\emph{arXiv preprint arXiv:2506.06913}} (\bibinfo{year}{2025}).
\newblock


\bibitem[Hao et~al\mbox{.}(2025)]%
        {hao2025oxygenrec}
\bibfield{author}{\bibinfo{person}{Xuegang Hao}, \bibinfo{person}{Ming Zhang}, \bibinfo{person}{Alex Li}, \bibinfo{person}{Xiangyu Qian}, \bibinfo{person}{Zhi Ma}, \bibinfo{person}{Yanlong Zang}, \bibinfo{person}{Shijie Yang}, \bibinfo{person}{Zhongxuan Han}, \bibinfo{person}{Xiaolong Ma}, \bibinfo{person}{Jinguang Liu}, {et~al\mbox{.}}} \bibinfo{year}{2025}\natexlab{}.
\newblock \showarticletitle{OxygenREC: An Instruction-Following Generative Framework for E-commerce Recommendation}.
\newblock \bibinfo{journal}{\emph{arXiv preprint arXiv:2512.22386}} (\bibinfo{year}{2025}).
\newblock


\bibitem[Kerr et~al\mbox{.}(2017)]%
        {Kerr2017cutlass}
\bibfield{author}{\bibinfo{person}{Andrew Kerr}, \bibinfo{person}{Duane Merrill}, \bibinfo{person}{Julien Demouth}, {and} \bibinfo{person}{John Tran}.} \bibinfo{year}{2017}\natexlab{}.
\newblock \bibinfo{title}{CUTLASS: Fast Linear Algebra in CUDA C++}.
\newblock \bibinfo{howpublished}{NVIDIA Developer Blog}.
\newblock
\urldef\tempurl%
\url{https://developer.nvidia.com/blog/cutlass-linear-algebra-cuda/}
\showURL{%
\tempurl}


\bibitem[Li et~al\mbox{.}(2025)]%
        {li2025cross}
\bibfield{author}{\bibinfo{person}{Xiuze Li}, \bibinfo{person}{Zhenhua Huang}, \bibinfo{person}{Zhengyang Wu}, \bibinfo{person}{Changdong Wang}, {and} \bibinfo{person}{Yunwen Chen}.} \bibinfo{year}{2025}\natexlab{}.
\newblock \showarticletitle{Cross-domain recommendation via knowledge distillation}.
\newblock \bibinfo{journal}{\emph{Knowledge-Based Systems}}  \bibinfo{volume}{311} (\bibinfo{year}{2025}), \bibinfo{pages}{113112}.
\newblock


\bibitem[Liu et~al\mbox{.}(2023)]%
        {liu2023visual}
\bibfield{author}{\bibinfo{person}{Haotian Liu}, \bibinfo{person}{Chunyuan Li}, \bibinfo{person}{Qingyang Wu}, {and} \bibinfo{person}{Yong~Jae Lee}.} \bibinfo{year}{2023}\natexlab{}.
\newblock \showarticletitle{Visual instruction tuning}.
\newblock \bibinfo{journal}{\emph{Advances in neural information processing systems}}  \bibinfo{volume}{36} (\bibinfo{year}{2023}), \bibinfo{pages}{34892--34916}.
\newblock


\bibitem[Liu et~al\mbox{.}(2025)]%
        {liu2025onerec}
\bibfield{author}{\bibinfo{person}{Zhanyu Liu}, \bibinfo{person}{Shiyao Wang}, \bibinfo{person}{Xingmei Wang}, \bibinfo{person}{Rongzhou Zhang}, \bibinfo{person}{Jiaxin Deng}, \bibinfo{person}{Honghui Bao}, \bibinfo{person}{Jinghao Zhang}, \bibinfo{person}{Wuchao Li}, \bibinfo{person}{Pengfei Zheng}, \bibinfo{person}{Xiangyu Wu}, {et~al\mbox{.}}} \bibinfo{year}{2025}\natexlab{}.
\newblock \showarticletitle{Onerec-think: In-text reasoning for generative recommendation}.
\newblock \bibinfo{journal}{\emph{arXiv preprint arXiv:2510.11639}} (\bibinfo{year}{2025}).
\newblock


\bibitem[Lu et~al\mbox{.}(2025)]%
        {lu2025large}
\bibfield{author}{\bibinfo{person}{Hui Lu}, \bibinfo{person}{Zheng Chai}, \bibinfo{person}{Yuchao Zheng}, \bibinfo{person}{Zhe Chen}, \bibinfo{person}{Deping Xie}, \bibinfo{person}{Peng Xu}, \bibinfo{person}{Xun Zhou}, {and} \bibinfo{person}{Di Wu}.} \bibinfo{year}{2025}\natexlab{}.
\newblock \showarticletitle{Large Memory Network for Recommendation}. In \bibinfo{booktitle}{\emph{Companion Proceedings of the ACM on Web Conference 2025}}. \bibinfo{pages}{1162--1166}.
\newblock


\bibitem[Ma et~al\mbox{.}(2018)]%
        {ma2018modeling}
\bibfield{author}{\bibinfo{person}{Jiaqi Ma}, \bibinfo{person}{Zhe Zhao}, \bibinfo{person}{Xinyang Yi}, \bibinfo{person}{Jilin Chen}, \bibinfo{person}{Lichan Hong}, {and} \bibinfo{person}{Ed~H Chi}.} \bibinfo{year}{2018}\natexlab{}.
\newblock \showarticletitle{Modeling task relationships in multi-task learning with multi-gate mixture-of-experts}. In \bibinfo{booktitle}{\emph{Proceedings of the 24th ACM SIGKDD international conference on knowledge discovery \& data mining}}. \bibinfo{pages}{1930--1939}.
\newblock


\bibitem[Tang et~al\mbox{.}(2020)]%
        {tang2020progressive}
\bibfield{author}{\bibinfo{person}{Hongyan Tang}, \bibinfo{person}{Junning Liu}, \bibinfo{person}{Ming Zhao}, {and} \bibinfo{person}{Xudong Gong}.} \bibinfo{year}{2020}\natexlab{}.
\newblock \showarticletitle{Progressive layered extraction (ple): A novel multi-task learning (mtl) model for personalized recommendations}. In \bibinfo{booktitle}{\emph{Proceedings of the 14th ACM conference on recommender systems}}. \bibinfo{pages}{269--278}.
\newblock


\bibitem[Wang et~al\mbox{.}(2026)]%
        {wang2026onelive}
\bibfield{author}{\bibinfo{person}{Shen Wang}, \bibinfo{person}{Yusheng Huang}, \bibinfo{person}{Ruochen Yang}, \bibinfo{person}{Shuang Wen}, \bibinfo{person}{Pengbo Xu}, \bibinfo{person}{Jiangxia Cao}, \bibinfo{person}{Yueyang Liu}, \bibinfo{person}{Kuo Cai}, \bibinfo{person}{Chengcheng Guo}, \bibinfo{person}{Shiyao Wang}, {et~al\mbox{.}}} \bibinfo{year}{2026}\natexlab{}.
\newblock \showarticletitle{OneLive: Dynamically Unified Generative Framework for Live-Streaming Recommendation}.
\newblock \bibinfo{journal}{\emph{arXiv preprint arXiv:2602.08612}} (\bibinfo{year}{2026}).
\newblock


\bibitem[Wei et~al\mbox{.}(2025)]%
        {wei2025oneloc}
\bibfield{author}{\bibinfo{person}{Zhipeng Wei}, \bibinfo{person}{Kuo Cai}, \bibinfo{person}{Junda She}, \bibinfo{person}{Jie Chen}, \bibinfo{person}{Minghao Chen}, \bibinfo{person}{Yang Zeng}, \bibinfo{person}{Qiang Luo}, \bibinfo{person}{Wencong Zeng}, \bibinfo{person}{Ruiming Tang}, \bibinfo{person}{Kun Gai}, {et~al\mbox{.}}} \bibinfo{year}{2025}\natexlab{}.
\newblock \showarticletitle{Oneloc: Geo-aware generative recommender systems for local life service}.
\newblock \bibinfo{journal}{\emph{arXiv preprint arXiv:2508.14646}} (\bibinfo{year}{2025}).
\newblock


\bibitem[Xia et~al\mbox{.}(2022)]%
        {xia2022device}
\bibfield{author}{\bibinfo{person}{Xin Xia}, \bibinfo{person}{Hongzhi Yin}, \bibinfo{person}{Junliang Yu}, \bibinfo{person}{Qinyong Wang}, \bibinfo{person}{Guandong Xu}, {and} \bibinfo{person}{Quoc Viet~Hung Nguyen}.} \bibinfo{year}{2022}\natexlab{}.
\newblock \showarticletitle{On-device next-item recommendation with self-supervised knowledge distillation}. In \bibinfo{booktitle}{\emph{Proceedings of the 45th International ACM SIGIR Conference on Research and Development in Information Retrieval}}. \bibinfo{pages}{546--555}.
\newblock


\bibitem[Zhou et~al\mbox{.}(2020)]%
        {zhou2020can}
\bibfield{author}{\bibinfo{person}{Guorui Zhou}, \bibinfo{person}{Weijie Bian}, \bibinfo{person}{Kailun Wu}, \bibinfo{person}{Lejian Ren}, \bibinfo{person}{Qi Pi}, \bibinfo{person}{Yujing Zhang}, \bibinfo{person}{Can Xiao}, \bibinfo{person}{Xiang-Rong Sheng}, \bibinfo{person}{Na Mou}, \bibinfo{person}{Xinchen Luo}, {et~al\mbox{.}}} \bibinfo{year}{2020}\natexlab{}.
\newblock \showarticletitle{CAN: revisiting feature co-action for click-through rate prediction}.
\newblock \bibinfo{journal}{\emph{arXiv preprint arXiv:2011.05625}} (\bibinfo{year}{2020}).
\newblock


\bibitem[Zhou et~al\mbox{.}(2025)]%
        {zhou2025onerec}
\bibfield{author}{\bibinfo{person}{Guorui Zhou}, \bibinfo{person}{Hengrui Hu}, \bibinfo{person}{Hongtao Cheng}, \bibinfo{person}{Huanjie Wang}, \bibinfo{person}{Jiaxin Deng}, \bibinfo{person}{Jinghao Zhang}, \bibinfo{person}{Kuo Cai}, \bibinfo{person}{Lejian Ren}, \bibinfo{person}{Lu Ren}, \bibinfo{person}{Liao Yu}, {et~al\mbox{.}}} \bibinfo{year}{2025}\natexlab{}.
\newblock \showarticletitle{Onerec-v2 technical report}.
\newblock \bibinfo{journal}{\emph{arXiv preprint arXiv:2508.20900}} (\bibinfo{year}{2025}).
\newblock


\bibitem[Zhou et~al\mbox{.}(2019)]%
        {zhou2019deep}
\bibfield{author}{\bibinfo{person}{Guorui Zhou}, \bibinfo{person}{Na Mou}, \bibinfo{person}{Ying Fan}, \bibinfo{person}{Qi Pi}, \bibinfo{person}{Weijie Bian}, \bibinfo{person}{Chang Zhou}, \bibinfo{person}{Xiaoqiang Zhu}, {and} \bibinfo{person}{Kun Gai}.} \bibinfo{year}{2019}\natexlab{}.
\newblock \showarticletitle{Deep interest evolution network for click-through rate prediction}. In \bibinfo{booktitle}{\emph{Proceedings of the AAAI conference on artificial intelligence}}, Vol.~\bibinfo{volume}{33}. \bibinfo{pages}{5941--5948}.
\newblock


\bibitem[Zhou et~al\mbox{.}(2018)]%
        {zhou2018deep}
\bibfield{author}{\bibinfo{person}{Guorui Zhou}, \bibinfo{person}{Xiaoqiang Zhu}, \bibinfo{person}{Chenru Song}, \bibinfo{person}{Ying Fan}, \bibinfo{person}{Han Zhu}, \bibinfo{person}{Xiao Ma}, \bibinfo{person}{Yanghui Yan}, \bibinfo{person}{Junqi Jin}, \bibinfo{person}{Han Li}, {and} \bibinfo{person}{Kun Gai}.} \bibinfo{year}{2018}\natexlab{}.
\newblock \showarticletitle{Deep interest network for click-through rate prediction}. In \bibinfo{booktitle}{\emph{Proceedings of the 24th ACM SIGKDD international conference on knowledge discovery \& data mining}}. \bibinfo{pages}{1059--1068}.
\newblock


\bibitem[Zhu et~al\mbox{.}(2025)]%
        {zhu2025rankmixer}
\bibfield{author}{\bibinfo{person}{Jie Zhu}, \bibinfo{person}{Zhifang Fan}, \bibinfo{person}{Xiaoxie Zhu}, \bibinfo{person}{Yuchen Jiang}, \bibinfo{person}{Hangyu Wang}, \bibinfo{person}{Xintian Han}, \bibinfo{person}{Haoran Ding}, \bibinfo{person}{Xinmin Wang}, \bibinfo{person}{Wenlin Zhao}, \bibinfo{person}{Zhen Gong}, {et~al\mbox{.}}} \bibinfo{year}{2025}\natexlab{}.
\newblock \showarticletitle{Rankmixer: Scaling up ranking models in industrial recommenders}. In \bibinfo{booktitle}{\emph{Proceedings of the 34th ACM International Conference on Information and Knowledge Management}}. \bibinfo{pages}{6309--6316}.
\newblock


\end{thebibliography}

\appendix
\section{Related Works}
\label{app:related_works}
Existing acceleration solutions for ranking models can be broadly categorized into three lines of research:
knowledge distillation, lightweight model design, and model–engine co-optimization.

\noindent\textbf{\textit{Knowledge distillation.}}
Knowledge distillation serves as a widely-used paradigm in RS, which transfers knowledge from large, cumbersome teacher models (e.g., LLMs or fine-ranking models) to smaller and more efficient student models (e.g., on-device models, pre-ranking or coarse-ranking models).
For instance, \citet{xia2022device} employ knowledge distillation to compress large-scale RS models into compact variants for efficient on-device deployment;
\citet{du2025active} distill world knowledge from LLMs into lightweight RS models;
and \citet{li2025cross} enable cross-domain recommendation via cross-domain knowledge transfer.
However, these methods primarily focus on knowledge transferring from teacher to student model rather than directly optimizing the computation of existing models.

\noindent\textbf{\textit{Lightweight model design.}}
The core philosophy of this line of research is to enhance the intrinsic computational efficiency of models at the architectural design stage.
For example, the Q-Former architecture \cite{liu2023visual} has been widely adopted in extreme long user sequence modeling for efficient historical information extraction \cite{chai2025longer,hao2025oxygenrec};
the sparse MoE structure is applied to high-order feature interaction \cite{zhu2025rankmixer}, improving model capacity while maintaining affordable inference overhead via sparse routing;
memory networks are introduced for user interest modeling \cite{lu2025large}, replacing online inference with efficient memory lookup.
Nevertheless, the above methods rarely focus on accelerating MatMul operations in the deployed ranking models, which serves as the key motivation for this work.

\noindent\textbf{\textit{Model–engine co-optimization.}}
Caching represents a widely adopted strategy for optimizing inference efficiency in RS.
For instance, recent advances in generative RS models \cite{deng2025onerec,zhou2025onerec,wang2026onelive,wei2025oneloc} and long user-sequence RS models \cite{chai2025longer} employ KV-Cache to store the key-value states of tokenized user-side features, avoiding redundant recomputation in consecutive inference steps.
Offline and nearline caching constitutes another industrial mainstream practice: through full pre-computation and frequent incremental updates, user and item representations generated from computationally expensive models (e.g., LLMs) can be served at reasonable overhead \cite{liu2025onerec,hao2025oxygenrec}.
In addition, vector hashing and quantization are widely adopted to enable efficient embedding indexing and similarity retrieval \cite{bin2025real,chen2021end}.
Techniques such as parameter quantization, model pruning, and kernel fusion are also highly effective and integrated into industrial-grade inference engines.


\section{Computational Complexity Analysis}
\subsection{Analisis for UOI}
\label{app:flops_uoi}
In this section, we present a brief FLOPs analysis of cross-attention under single-head attention. The number of query vectors per candidate item is set to 1 for simplicity (i.e., query count $I=1$). Let $B$ denote the batch size (number of candidate items), $L$ the user historical sequence length, and $d$ the hidden embedding dimension.

In VanI, the user sequence is replicated $B$ times across the batch dimension, so the key/value projections are redundantly executed $B$ times.
The dominant computational cost of cross-attention is:
\[
\text{FLOPs}_{\text{vanilla}} \approx B d^2(1 + 2L).
\]

By contrast, our User-Side One-Shot Inference (UOI) computes the key/value projections on the original user sequence \emph{once} by avoiding batch-wise replication.
The corresponding cost is:
\[
\text{FLOPs}_{\text{UOI}} \approx B d^2 + 2L d^2.
\]

The FLOPs ratio between UOI and VanI is:
\[
\frac{\text{FLOPs}_{\text{UOI}}}{\text{FLOPs}_{\text{vanilla}}}
\approx \frac{B + 2L}{B(1 + 2L)} \ll 1,
\]

As $L\to\infty$, the FLOPs ratio of UOI to vanilla inference converges to $1/B$;
as $B\to\infty$, it approaches $1/(1+2L)$, both being far smaller than $1$ in practice.

\subsection{Analisis for MaRI}
\label{app:flops_mari}
Traditional MatMul computes $\mathbf{X^{\text{U\_Tiled}}}\mathbf{W}$ directly, leading to:
\begin{equation}
\text{FLOPs}_{\text{ori}} = 2 \cdot B \cdot D \cdot d = 2Bd(D_u + D_i + D_c).
\end{equation}
The core computational overhead stems from the tiled user features which results in $B \cdot D_u \cdot d$ redundant multiplications for user-side computation.

The structural re-parameterizated $\text{MatMul}_{\text{MaRI}}$ decomposes the original MatMul into three MatMul computations:
\begin{itemize}
    \item User-side: $\mathbf{X}_{\text{user}}\mathbf{W}_{\text{user}}$ with $\text{FLOPs}_u = 2 \cdot 1 \cdot D_u \cdot d$;
    \item Item-side: $\mathbf{X}_{\text{item}}\mathbf{W}_{\text{item}}$ with $\text{FLOPs}_i = 2 \cdot B \cdot D_i \cdot d$;
    \item Cross-side: $\mathbf{X}_{\text{cross}}\mathbf{W}_{\text{cross}}$ with $\text{FLOPs}_c = 2 \cdot B \cdot D_c \cdot d$.
\end{itemize}

The total FLOPs for $\text{MatMul}_{\text{MaRI}}$ is:
\begin{equation}
\text{FLOPs}_{\text{MaRI}} = 2d\left[D_u + B(D_i + D_c)\right].
\end{equation}

MaRI reduces FLOPs by avoiding redundant $B$-fold user-side MatMul, achieving an absolute saving of $\Delta\text{FLOPs}=2dD_u(B-1)$ and a relative saving ratio of $\text{Ratio}_{\text{save}}\approx D_u/(D_u+D_i+D_c)$ (since $B\gg1$ in industrial RS).
This decomposition introduces no extra FLOPs, and the savings grow more significant as the proportion of user feature dimension $D_u$ increases.

\section{The Graph Coloring Algorithm (GCA)}
We present the detailed pseudocode of our GCA in Algorithm~\ref{alg:gca} for detecting MaRI-optimizable nodes.
DFS instead of BFS is applied for proper color propagation. Traverse puring is required to avoid repeated traversals and is not demonstrated in the pseudocode.
\vspace{10pt}
\begin{algorithm}[tbh!]
  \caption{Graph Coloring Algorithm for the Detection of MaRI-Optimizable MatMul Nodes}
  \label{alg:gca}
  \begin{algorithmic}[1]
    \REQUIRE 
    $\mathcal{G}$: Ranking model computation graph (feature/operation nodes only);
    $\mathcal{F}_{\text{user}}/\mathcal{F}_{\text{item}}/\mathcal{F}_{\text{cross}}$: Feature node sets.
  \ENSURE 
    $\mathcal{M}_{\text{opt}}$: MaRI-optimizable MatMul nodes
  
  \STATE \textbf{1. Initialization}
  \FORALL{$v \in \mathcal{G}$}
      \STATE $\text{Color}(v) = \begin{cases}
          \text{Yellow}, & v \in \mathcal{F}_{\text{user}} \\
          \text{Blue}, & v \in \mathcal{F}_{\text{item}} \cup \mathcal{F}_{\text{cross}} \\
          \text{Uncolored}, & \text{otherwise}
      \end{cases}$
  \ENDFOR
  
  \STATE \textbf{2. DFS-based Color Propagation}
  \STATE $\mathcal{S} \leftarrow$ Stack of all colored nodes.
  \WHILE{$\mathcal{S} \neq \emptyset$}
      \STATE $u \leftarrow \text{Pop}(\mathcal{S})$
      \FORALL{downstream neighbor $v$ of $u$}
          \STATE $\text{updated} \leftarrow \text{False}$
          \IF{$\text{Color}(u)=\text{Blue}$ and $\text{Color}(v)\neq\text{Blue}$}
            \STATE $\text{updated} \leftarrow \text{True}$; $\text{Color}(v)\leftarrow\text{Blue}$
          \ELSIF{$\text{Color}(u)=\text{Yellow}$ and $\text{Color}(v)=\text{Uncolored}$}
            \STATE $\text{updated} \leftarrow \text{True}$; $\text{Color}(v)\leftarrow\text{Yellow}$
          \ENDIF
          \STATE $\text{Push}(\mathcal{S}, v)$ if $\text{updated}$
      \ENDFOR
  \ENDWHILE
  
  \STATE \textbf{3. Optimizable MatMul Detection}
  \STATE $\mathcal{M}_{\text{opt}} \leftarrow \emptyset$
  \FORALL{``Concat'' node $c \in \mathcal{G}$}
      \STATE $\mathcal{I}_c$: Input nodes of $c$
      \IF{$\mathcal{I}_c$ contains both Yellow and Blue nodes}
          \STATE $\mathcal{N}_{\text{matmul}} \leftarrow$ All MatMul nodes reachable from $c$ via paths with only non-computational nodes
          \STATE $\mathcal{M}_{\text{opt}} \leftarrow \mathcal{M}_{\text{opt}} \cup \mathcal{N}_{\text{matmul}}$
      \ENDIF
  \ENDFOR
  
  \RETURN $\mathcal{M}_{\text{opt}}$
  \end{algorithmic}

\end{algorithm}

\section{Detailed Experience Results}
\label{app:exp}
In this section, we provide additional experimental results to validate the effectiveness of our methods.

In Table~\ref{tab:exp_comprehensive}, we present comprehensive offline simulation results under varying configurations, demonstrating a consistent superiority of our MaRI across different settings.
In Table~\ref{tab:chunk_size_performance}, we present the degradation of the fragmented MaRI under different chunk sizes with fixed $D_{user}=4000$, $D_{item}=1000$, and $D_{hidden}=256$.

\begin{table}[h]
  \centering
  \caption{Comprehensive Offline Simulation Results: FLOPs and Runtime Speedup Under Different Parameter Configurations}
  \label{tab:exp_comprehensive}
  \resizebox{\linewidth}{!}{
  \begin{tabular}{ccccc}
    \toprule
    \textbf{Setting} & \makecell{\textbf{Variable}\\ \textbf{Value}} & \makecell{\textbf{Theoretical FLOPs}\\\textbf{Speedup}} & \makecell{\textbf{Time Speedup}\\ \textbf{($\pm$Std)}} \\
    \midrule
    \multirow{7}{*}{\makecell{Varying $B$\\(Fixed $D_{user}=4000$,\\ $D_{item}=1000$,\\ $D_{hidden}=512$)}} 
    & 100 & 2.94 & 0.97$\pm$0.018 \\
    & 500 & 2.99 & 1.12$\pm$0.005 \\
    & 1000 & 2.99 & \textbf{1.14$\pm$0.006} \\
    & 2000 & 3.00 & 1.10$\pm$0.006 \\
    & 5000 & 3.00 & 1.10$\pm$0.011 \\
    & 8000 & 3.00 & 1.11$\pm$0.007 \\
    & 10000 & 3.00 & 1.10$\pm$0.002 \\
    \midrule
    \multirow{6}{*}{\makecell{Varying $D_{\text{user}}$\\(Fixed $B=2000$, \\$D_{item}=1000$,\\ $D_{hidden}=512$)}} 
    & 500 & 1.50 & 1.02$\pm$0.007 \\
    & 1000 & 2.00 & 1.05$\pm$0.006 \\
    & 2000 & 3.00 & 1.10$\pm$0.004 \\
    & 5000 & 5.99 & 1.27$\pm$0.010 \\
    & 8000 & 8.96 & 1.43$\pm$0.026 \\
    & 10000 & 10.95 & \textbf{1.58$\pm$0.023} \\
    \midrule
    \multirow{6}{*}{\makecell{Varying $D_{\text{item/cross}}$\\ (Fixed $B=2000$,\\ $D_{user}=4000$,\\ $D_{hidden}=512$)}} 
    & 500 & 8.96 & \textbf{1.61$\pm$0.008} \\
    & 1000 & 4.99 & 1.22$\pm$0.005 \\
    & 2000 & 3.00 & 1.11$\pm$0.003 \\
    & 5000 & 1.80 & 1.05$\pm$0.003 \\
    & 8000 & 1.50 & 1.04$\pm$0.006 \\
    & 10000 & 1.40 & 1.04$\pm$0.004 \\
    \midrule
    \multirow{4}{*}{\makecell{Varying $D_{hidden}$\\(Fixed $B=2000$, \\ $D_{user}=4000$,\\ $D_{item}=1000$)}} 
    & 128 & 4.99 & 1.10$\pm$0.007 \\
    & 512 & 4.99 & 1.21$\pm$0.008 \\
    & 1024 & 4.99 & 1.31$\pm$0.004 \\
    & 2048 & 4.99 & \textbf{1.58$\pm$0.006} \\
    \bottomrule
  \end{tabular}
  }
  \\
  \footnotesize
  \textit{Note}: Time speedup is calculated based on method's inference time $=\frac{T_{\text{Ori}}}{T_{\text{MaRI}}}$.
\end{table}

\begin{table}[b]
  \centering
  \caption{Performance of fragmented MaRI with Different Chunk Sizes}
  \resizebox{\linewidth}{!}{
  \begin{tabular}{lcc}
  \toprule
  \textbf{Chunk Size} & \makecell{\textbf{Degradation} \\ \textbf{vs. Original (\%)}} & \makecell{\textbf{Degradation} \\\textbf{vs. neat MaRI (\%)}} \\
  \midrule
  50  & 69.4 & 96.3 \\
  100 & 27.2 & 47.4 \\
  200 & 13.6 & 31.7 \\
  400 & 3.1  & 19.5 \\
  800 & -0.7 & 15.1 \\
  \bottomrule
  \end{tabular}
  }
  \\
  \footnotesize
  \textit{Note}: Degradation is calculated based on method's inference time $=\frac{T_{\text{fragmented}} - T_{\text{baseline}} }{T_{\text{baseline}}}$.
  \label{tab:chunk_size_performance}
  \end{table}

\end{document}